\documentclass[preprint,showpacs,showkeys]{revtex4}
\begin{document}

\title{Invariant conserved currents for gravity}

\author{Yuri N.~Obukhov}
\email{yo@thp.uni-koeln.de}
\affiliation{Institute for Theoretical Physics, University of Cologne, 
50923 K\"oln, Germany\,}
\altaffiliation[Also at: ]{Department of Theoretical Physics, Moscow 
State University, 117234 Moscow, Russia}
\author{Guillermo F.~Rubilar}
\email{grubilar@udec.cl}
\affiliation{Departamento de F{\'{\i}}sica, Universidad de Concepci\'on,
Casilla 160-C, Concepci\'on, Chile}

\begin{abstract}
We develop a general approach, based on the Lagrange-Noether machinery, to the 
definition of invariant conserved currents for gravity theories
with general coordinate and local Lorentz symmetries. In this framework, every
vector field $\xi$ on spacetime generates, in any dimension $n$, for any
Lagrangian of 
gravitational plus matter fields and for any (minimal or nonminimal) type of 
interaction, a current ${\cal J}[\xi]$ with the following properties: (1) the 
current $(n-1)$-form ${\cal J}[\xi]$ is constructed from the Lagrangian and the 
generalized field momenta, (2) it is conserved, $d{\cal J}[\xi] = 0$, when the 
field equations are satisfied, (3) ${\cal J}[\xi]= d\Pi[\xi]$ ``on shell", (4) 
the current ${\cal J}[\xi]$, the superpotential $\Pi[\xi]$, and the 
charge ${\cal Q}[\xi] = \int {\cal J}[\xi]$ are invariant under diffeomorphisms 
and the local Lorentz group. We present a compact derivation of the Noether
currents associated with diffeomorphisms and apply the general method to compute 
the total energy and angular momentum of exact solutions in several physically
interesting gravitational models.
\end{abstract}


\maketitle

\section{Introduction to Noether currents}

Conserved currents are related to symmetries of a physical 
model. In gravity theories based on the covariance principle, the action is 
diffeomorphism-invariant. Diffeomorphisms are generated by vector 
fields. Hence, vector fields should give rise to conserved currents.

Let $\xi$ be a vector field on an $n$-dimensional spacetime. Then, indeed, 
we can associate a conserved charge to it in diffeomorphism-invariant models. 
For example, take a symmetric energy-momentum tensor $T_j{}^i$ (which is 
covariantly conserved in such theories) and a {\it Killing} field $\xi = 
\xi^i\partial_i$ (that generates an isometry of the spacetime). Then
$j^i = \xi^jT_j{}^i$ is a conserved current, and a
corresponding charge is defined \cite{footnote}
as the integral $\int_S j^i\,\partial_i\rfloor\eta$ over an 
$(n-1)$-hypersurface $S$. Moreover, it is possible to construct a conserved
current $(n-1)$-form for any solution of a diffeomorphism-invariant model 
even when $\xi$ is not a Killing field \cite{Wald}. Such a current and the
corresponding charge are scalars under general coordinate transformations.
The Komar charges \cite{Komar} arise in this way.  

The situation becomes more complicated when, besides the diffeomorphism 
symmetry, the gravitational model is also invariant under the local Lorentz 
group $SO(1,n-1)$. This is the case for the gauge gravity models 
\cite{Trautman05},
for supergravity, for the so called first order formulation of standard 
General Relativity, and, in general, for the case when {\it spinor} matter 
is present. The problem of defining conserved quantities associated with a 
vector field was analysed \cite{Trautman05,Benn,FFFR99,JS98,A00a} for 
specific Lagrangians (usually, for the Hilbert-Einstein one) and for specific
types of vector fields (usually, for Killing or generalized Killing
ones). Moreover, some resulting conserved quantities were {\it not invariant}
under the local Lorentz group (e.g., in \cite{A00a}). 

In this letter, we present a {\it compact derivation of the invariant
conserved currents and charges for gravity models with diffeomorphism and local
Lorentz symmetries}. This is done in any dimension, for any Lagrangian of the 
gravitational field and of (minimally or nonminimally coupled) matter,
and for any vector field $\xi$.  We then apply our general results to
different interesting gravitational models.

It is worthwhile to stress that the subject of our studies are not all
currents (and the corresponding charges), but only the {\it Noether currents}.
We recall, that Noether \cite{Noether} demonstrated that a symmetry of the action
gives rise to a current, the divergence of which is a linear combination of the 
field (Euler-Lagrange) equations. She proved also the inverse theorem: if there 
is a current, the divergence of which is equal to a linear combination of the 
field equations, then the action is invariant under a certain symmetry. In this 
way, there is a one-to-one correspondence between the symmetries of the action 
and the currents with the property mentioned. Furthermore, if the fields satisfy 
the Euler-Lagrange equations (``on-shell", using the standard jargon), then 
such a current is conserved, i.e., its divergence vanishes. It is possible 
then to define the corresponding charge. Because of the direct ``off-shell"
relation of the Noether current with the symmetry of the action (that is, with 
the structure of the latter), the form of such current is determined by the
Lagrangian of the theory.

In addition, in any field-theoretic model there is an infinite number of 
{\it non-Noether currents}. In $n$ dimensions, such a current is given by
an arbitrary exact $(n - 1)$-form, i.e., $J^{\rm nN}:=dU$ where an $(n-2)$-form 
$U$ can be an arbitrary function of the fields. Obviously, $d J^{\rm nN} \equiv 
0$, and this fact holds always, independently of the structure of the Lagrangian
and of the field equations. We stress this point: The divergence of such a
current is {\it not} equal to a linear combination of the field equations.
Accordingly, using the inverse Noether theorem, such a current {\it does not
correspond to any symmetry of the action}. That is the reason why we call this a
non-Noether current. As a result, in contrast to the Noether currents, the
structure of non-Noether currents is completely ambiguous. One needs additional
assumptions in order to fix their structure, the corresponding general
discussion can be found in \cite{Barnich}, for example. 

It is clear that given a Noether current $J$ and a non-Noether current 
$J^{\rm nN}$, one can define a new quantity $J':=J + J^{\rm nN}$. This is
conserved ``on-shell". Accordingly, the new conserved charge is defined
as a sum of the Noether charge and a non-Noether charge. It is important 
to realize that the Noether charge and non-Noether charges are conserved 
{\it separately}. If we recall, in addition, that non-Noether currents 
and charges are completely arbitrary and not related either to the Lagrangian 
of a theory or to its dynamics, then a reasonable question arises: Why at 
all do we need to mix the two quantities, $J$ and $J^{\rm nN}$? One can, 
instead, consistently study {\it only} Noether currents, which are directly 
related to the symmetries of the action, and the structure of which is fixed 
by the crucial ``off-shell" relation of their divergence to the combination 
of the Euler-Lagrange equations. And, separately, one can consistently study
{\it only} non-Noether currents, which are not related to any symmetry and 
even to any Lagrangian as such, and the structure of which should
be fixed by totally different considerations. Since the study of the 
latter can be found elsewhere \cite{Barnich} (see also the recent
development in \cite{Barnich2}), here we concentrate on the {\it Noether
currents only}. 

We then demonstrate that the structure of the Noether current associated
with an arbitrary diffeomorphism $\xi$ is uniquely determined by the 
``off-shell" relation (\ref{dJF}). Furthermore, we show that the structure
of the corresponding superpotential $(n-2)$-form is also uniquely fixed 
by the ``off-shell" relation (\ref{Jd}). We stress this crucial point: 
Following the original Noether's derivations \cite{Noether}, we find a
precise form of the current and superpotential for every Lagrangian, which
is invariant under diffeomorphisms and the local Lorentz group, in terms
of the Lagrangian and its derivatives (\ref{HE}). There is no any ambiguity
in our construction, since from the very beginning we have put aside all the
ambiguously defined non-Noether currents. 

The only ``ambiguity" left in our approach is related to the fact that the
Lagrangian can be shifted by a total derivative (boundary) term. But this
is a controlled ``ambiguity" (and, hence, not an ambiguity at all): When
we change the Lagrangian in this way, we automatically change its derivatives
(\ref{HE}) with respect to the fields $\Phi$ and ``velocities" $d\Phi$. As a
result, we certainly obtain a new current for a transformed Lagrangian. But 
the form of the new current is again uniquely determined by the new (shifted 
by a boundary term) Lagrangian. The one-to-one correspondents between the 
Lagrangians and the Noether currents is thus guaranteed in this approach.  

Before we go ahead with the description of the results, it seems necessary
to make a comment concerning some misunderstandings about the Noether currents
and charges. A typical example can be found in \cite{Barnich}, where on page 4 
the authors write the following: ``... a Noether current associated to a gauge 
symmetry necessarily vanishes on-shell ... up to a divergence of an arbitrary 
superpotential". Furthermore, on page 5 they continue with: ``Note that the 
superpotential is completely {\it arbitrary} (the emphasis of the authors of 
\cite{Barnich})... . This implies in particular that the Noether charge ... 
is {\it undefined} (our emphasis) because it is given by the surface integral 
of an arbitrary $(n-2)$-form". These statements are misleading. 

To begin with, the Noether current certainly does not ``vanish on-shell". It
is closed on-shell. But the current $J$ itself is nonvanishing
and it defines a conserved charge $Q = \int_S J$ (when the fields satisfy certain
appropriate conditions at infinity, which is usually assumed). It is worthwhile
to remember that a charge $Q$ is, as such, a {\it volume} integral over an 
$(n-1)$-dimensional spatial hypersurface $S$. And in this sense, the crucial 
thing for the evaluation of the total charge $Q$ is the {\it physics inside} 
this volume $S$, and not mathematics at the boundary $\partial S$ (recall how a 
conserved electric charge is evaluated in classical electrodynamics, for example).

Furthermore, it is certainly well known (see, \cite{Barnich,Wald2}, for
example), that the ``on-shell conserved" currents for local symmetry groups are
``on-shell exact", that is, they are expressed in terms of superpotentials.
We also prove this explicitly here. However, this fact does not devaluate the
Noether construction in any sense. Indeed, suppose on-shell we have $J =
d\Pi_0$. Let us ask: To what extent the superpotential $(n-2)$-form is
arbitrary? To study this, we assume that besides $\Pi_0$ there is another 
form $\Pi_1$ which also satisfies $J = d\Pi_1$. 
Now, by a simple subtraction of the first equation from the second one, we
find that the difference $(\Pi_1 - \Pi_0)$ is closed: $d(\Pi_1 - \Pi_0) = 0$.
Then, using the results on the trivial cohomologies \cite{Barnich} we derive
that $\Pi_1 = \Pi_0 + dU$ where $U$ is an arbitrary $(n-3)$-form. Therefore, a
superpotential $\Pi$ for a Noether current $J = d\Pi$ of a gauge symmetry 
is indeed defined nonuniquely, but the corresponding ambiguity is just a shift
by an exact form, $\Pi \longrightarrow \Pi + dU$. It is easy to see that this 
``arbitrariness" is harmless. Indeed, starting from the charge defined as a
volume integral $Q = \int_S J$, after substituting $J = d\Pi$ and using 
the Stokes theorem, we reduce the charge to the surface integral $Q = 
\int_{\partial S}\Pi$. Now, what is the effect of a shift $\Pi \longrightarrow
\Pi + dU$ on the value of this integral? Well, there is no effect at all, 
because $\int_{\partial S}dU \equiv 0$ (use Stokes theorem and ``boundary of
a boundary is zero"). In other words, the statement of \cite{Barnich} that 
a Noether charge is ``undefined" because of the ambiguity of a superpotential 
is misleading and wrong. The Noether charge is {\it well defined} despite a 
certain arbitrariness in the choice of a superpotential.
 
Previously, we have demonstrated \cite{OR06b} that the Noether-Lagrange approach 
works perfectly well in the standard case of the Einstein(-Cartan) theory, and 
the computation of the total energy and angular momentum is in an agreement with 
results obtained by other methods (ADM mass, Hamiltonian surface integrals, and 
covariant phase space charges) \cite{Wald,FFFR99,JS98}. Here we apply this 
approach to other gravitational models, such as theories in lower and higher 
dimension, Brans-Dicke and higher derivative models.

\section{General formulation}

In the theories possessing local Lorentz invariance, the gravitational 
field is naturally described by the 1-forms of the coframe $\vartheta^\alpha$
and the Lorentz connection $\Gamma_\alpha{}^\beta$. The orthonormal coframe 
determines the lengths and angles on the spacetime manifold by introducing
the line element $ds^2 = g_{\alpha\beta}\,\vartheta^\alpha\otimes\vartheta
{}^\beta$, with the $n$-dimensional Minkowski tensor given by $g_{\alpha\beta}
:= {\rm diag}(1,-1,\cdots,-1)$. Using the gauge-theoretic language, we will
refer to the coframe and connection as the translational and rotational 
fields, respectively. In this framework, both the curvature 2-form, 
$R_\alpha{}^\beta = d\Gamma_\alpha{}^\beta + \Gamma_\gamma{}^\beta\wedge
\Gamma_\alpha{}^\gamma$, and the 2-form of torsion, $T^\alpha = D
\vartheta^\alpha$, are nontrivial, in general. 

Let us consider a Lagrangian $n$-form $V^{\rm tot} = V + L$ that describes 
the system of coupled gravitational, $\vartheta^\alpha, \Gamma_\alpha
{}^\beta$, and matter fields, $\psi^A$. The latter include scalars 
and spinors of arbitrary rank, belonging to some representation of the
Lorentz group. We assume them to be $0$-forms. The gravitational Lagrangian 
$V = V(\vartheta^\alpha, T^\alpha, R_\alpha{}^\beta)$ depends on the  
covariant geometrical objects: coframe, torsion and curvature. The material
Lagrangian $L = L(\psi^A, D\psi^A, \vartheta^\alpha, T^\alpha, R_\alpha
{}^\beta)$ is a function of the matter fields and their derivatives, but 
it may also depend on the coframe, torsion and curvature. By allowing this,
we include the possibility of nonminimal coupling of matter to gravity (like 
in the Brans-Dicke theory, for example) by means of Pauli-type terms in the
action. Moreover, some of the matter fields may play the role of nondynamical
Lagrange multipliers imposing various constraints. For example, the 
zero-torsion constraint $T^\alpha =0$ can be introduced by means of the 
term $\psi_\alpha\wedge T^\alpha$ with the help of a Lagrange multiplier 
$(n-2)$-form $\psi_\alpha$. The resulting dynamical setting of theories 
with local Lorentz symmetry generalizes the previous studies 
\cite{FFFR99,JS98}.

It is convenient to collect all the fields (gravitational and matter) 
into a single multi-component field: $\Phi^I = \left(\vartheta^\alpha, 
\Gamma_\alpha{}^\beta, \psi^A\right)$. The collective index $I$ runs over
the three sectors: $\alpha$ (``translational"), $[\alpha\beta]$
(``rotational", labeled by antisymmetrized pairs of indices), and $A$ 
(matter). The derivatives of $V^{\rm tot}(\Phi^I, d\Phi^I)$ w.r.t. the
generalized ``velocities" and w.r.t. the fields introduce the field 
momenta and the ``potential energy" terms by
\begin{equation}
H_I := -\,{\frac {\partial V^{\rm tot}}{\partial d\Phi^I}},\qquad
E_I := {\frac {\partial V^{\rm tot}}{\partial \Phi^I}},\label{HE}
\end{equation}
respectively. A total variation of the Lagrangian is then 
\begin{equation}
\delta V^{\rm tot} = \delta\Phi^I\wedge {\cal F}_I - 
d(\delta\Phi^I\wedge H_I),\label{deltaV1}
\end{equation}
where we introduced the variational derivative
\begin{equation}
{\cal F}_I := {\frac {\delta V^{\rm tot}}{\delta \Phi^I}} =
(-1)^{p(I)}DH_I + E_I. 
\end{equation}
Here $p(I)$ denotes the rank (in the exterior sense) of the 
corresponding sector of the collective field ($p=1$
for the coframe and connection and $p=0$ for the matter field).
The field equations for the coupled gravitational and 
matter fields read ${\cal F}_I =0$.

We assume that the action of the theory is invariant under 
diffeomorphism and local Lorentz transformations. The total 
{\it infinitesimal symmetry variation} of the collective 
field consists of two terms:
\begin{equation}
\delta\Phi^I = \varsigma{\cal L}_{\{\xi,\varepsilon\}}\Phi^I:=
\delta_{(\varsigma\xi)}\Phi^I +\delta_{(\varsigma\varepsilon)}
\Phi^I . \label{defL}
\end{equation}
Here $\varsigma$ is an infinitesimal constant parameter. The 
term $\delta_{(\varsigma\xi)}\Phi^I=\varsigma\ell_\xi\Phi^I$ comes
from a diffeomorphism generated by a vector field $\xi$, where
$\ell_\xi$ is the Lie derivative along the latter. The second term
describes a local Lorentz transformation $\delta_{(\varsigma
\varepsilon)}\Phi^I = \varsigma\left[\varepsilon^\alpha{}_\beta
(\rho^\beta{}_\alpha)^I{}_J\,\Phi^J - (\sigma^\beta{}_\alpha)^I
\,d\varepsilon^\alpha{}_\beta\right]$. Here $\rho^I{}_J$ are the
Lorentz generators, and $\sigma^I$ is nontrivial only in the 
rotational sector (connection), in which it 
is equal to the identity matrix. 
It is convenient to introduce a special notation, ${\cal
L}_{\{\xi,\varepsilon\}}$, for the total variation, as we did in 
the last equality of (\ref{defL}). 

The condition of the invariance of the theory under a general 
variation (\ref{defL}) is read off directly from (\ref{deltaV1}):
\begin{equation}\label{deltaV2}
d(\xi\rfloor V^{\rm tot}) = 
({\cal L}_{\{\xi,\varepsilon\}}\Phi^I)\wedge{\cal F}_I -
d\left[({\cal L}_{\{\xi,\varepsilon\}}\Phi^I)\wedge H_I\right]. 
\end{equation}
Introducing the current $(n-1)$-form 
\begin{equation}
J[\xi,\varepsilon] := \xi\rfloor V^{\rm tot} + 
({\cal L}_{\{\xi,\varepsilon\}}\Phi^I)\wedge H_I,\label{Jdef}
\end{equation}
we see from (\ref{deltaV2}) that
\begin{equation}
dJ[\xi,\varepsilon] = ({\cal L}_{\{\xi,\varepsilon\}}\Phi^I)
\wedge{\cal F}_I.\label{dJF}
\end{equation}
Hence, this current is conserved, $dJ[\xi,\varepsilon] = 0$, 
for any $\xi$ and $\varepsilon^\alpha{}_\beta$, when the field
equations, ${\cal F}_I =0$, are satisfied.  

Using (\ref{Jdef}), (\ref{defL}) and the Noether identities of the 
diffeomorphism symmetry \cite{OR06b}, we rewrite the current as
\begin{equation}
J[\xi,\varepsilon] = d\Pi[\xi,\varepsilon]
+ \Xi^I[\xi,\varepsilon]\wedge {\cal F}_I.\label{Jd}
\end{equation}
Here we denoted $\Pi[\xi,\varepsilon] := \Xi^I[\xi,\varepsilon]
\wedge H_I$ and 
\begin{equation}
\Xi^I[\xi,\varepsilon] := \xi\rfloor\Phi^I - \varepsilon^\alpha
{}_\beta(\sigma^\beta{}_\alpha)^I.\label{Xi}
\end{equation}
On the solutions of the field equations, the charge is an integral over an
$(n-2)$-boundary:
\begin{equation}
Q[\xi,\varepsilon] = \int_S J[\xi,\varepsilon] = 
\int_{\partial S} \Pi[\xi,\varepsilon].\label{charge}
\end{equation}

The functions $\varepsilon^\alpha{}_\beta$ parametrize the {\it
family} of conserved currents (\ref{Jdef}) and charges (\ref{charge})
associated with a vector field $\xi$. In order to select {\it invariant}
charges, we will have to specialize to a particular
choice of $\varepsilon$. The trivial choice $\varepsilon^\alpha{}_\beta=0$ 
yields a {\it noninvariant} current and charge (an explicit example 
was obtained recently in \cite{A00a}). Indeed, then ${\cal L}_{\{\xi,
\varepsilon\}}\Phi^I = \ell_\xi\Phi^I$, and the last term in (\ref{Jdef})
is not Lorentz invariant, since the Lie derivative $\ell_\xi$ is not 
covariant under local Lorentz transformations.
 
The situation is improved if we make ${\cal L}_{\{\xi,\varepsilon\}}$ 
a covariant operator by an appropriate choice of $\varepsilon$. This 
is always possible to do, although not uniquely. The choice 
\begin{equation}
\varepsilon_{\alpha\beta}=-\Theta_{\alpha\beta} := -e_{[\alpha}
\rfloor\ell_\xi\vartheta_ {\beta]}\label{Theta}
\end{equation}
is in a certain sense \textit{minimal}. The Lie derivative of the coframe 
can be decomposed as $\ell_\xi\vartheta^\alpha=\left(S_\beta{}^\alpha + 
\Theta_\beta{}^\alpha\right) \vartheta^\beta$, where the symmetric 
$S_{\alpha\beta}= e_{(\alpha}\rfloor\ell_\xi\vartheta_{\beta)}\equiv 
h^i_\alpha h^j_\beta\,\ell_\xi g_{ij}/2$ and the antisymmetric $\Theta_\beta
{}^\alpha$ is given by (\ref{Theta}). We can immediately verify that 
$S_\beta{}^\alpha$ is a tensor under local Lorentz transformations. Then 
we simply move $\Theta_\beta{}^\alpha \vartheta^\beta$  to the l.h.s. and
define a ``generalized Lie derivative" of the coframe as ${\cal L}_\xi
\vartheta^\alpha := \ell_\xi\vartheta^\alpha - \Theta_\beta{}^\alpha
\vartheta^\beta$. By construction, it is {\it covariant}, and, moreover,
${\cal L}_\xi\vartheta^\alpha = {\cal L}_{\{\xi,\varepsilon = - \Theta\}}
\vartheta^\alpha$. Thus, the choice (\ref{Theta}) is minimal in the sense
that it provides a covariant generalization of the Lie derivative \cite{Lie}
of the coframe without any additional variables and constants, using just
the coframe itself. Inserting (\ref{Theta}) into (\ref{Xi}), we find 
$\Xi_\alpha{}^\beta = \xi\rfloor\Gamma_\alpha{}^\beta + \Theta_\alpha
{}^\beta$.

Our formalism thus shows that for a gravity model with diffeomorphism and 
local Lorentz symmetries, any vector field $\xi$ generates an {\it invariant 
current} (${\cal L}_\xi := {\cal L}_{\{\xi,\varepsilon = - \Theta\}}$):
\begin{equation}
{\cal J}[\xi] = \xi\rfloor V^{\rm tot}
+ {\cal L}_\xi\vartheta^\alpha
\wedge H_\alpha + {\cal L}_\xi\Gamma_\alpha{}^\beta\wedge H^\alpha{}_\beta 
+ {\cal L}_\xi\psi^A H_A.\label{Jdef1}
\end{equation}
We expand here the condensed notation in order to show how the 
gravitational and matter fields appear in the final formulas. The 
definition (\ref{HE}) reads: $H_\alpha = - \partial V^{\rm tot}/\partial
T^\alpha$, $H^\alpha{}_\beta = - \partial V^{\rm tot}/\partial R_\alpha
{}^\beta$, and $H_A = -\partial V^{\rm tot}/\partial D\psi^A$. The 
current (\ref{Jdef1}) and its derivative satisfy:
\begin{eqnarray}
{\cal J}[\xi] = d\Pi[\xi] + \,\xi^\alpha{\cal F}_\alpha
+ \Xi_\alpha{}^\beta {\cal F}^\alpha{}_\beta, \label{Jd2}\\
d{\cal J}[\xi] = {\cal L}_\xi\vartheta^\alpha\wedge {\cal F}_\alpha
+ {\cal L}_\xi\Gamma_\alpha{}^\beta\wedge {\cal F}^\alpha{}_\beta
+ {\cal L}_\xi\psi^A {\cal F}_A.\label{dcurr}
\end{eqnarray}
Here $\Pi[\xi] := \xi^\alpha H_\alpha + \Xi_\alpha{}^\beta H^\alpha
{}_\beta$. For the solutions of the field equations, the {\it invariant 
charge} (\ref{charge}) then reads
\begin{equation}
{\cal Q}[\xi] = \int_{\partial S}\left(\xi^\alpha H_\alpha
+ \Xi_\alpha{}^\beta H^\alpha{}_\beta\right),\label{calq}
\end{equation} 
with the $(n-2)$-dimensional boundary $\partial S$ of a spacelike 
$(n-1)$-hypersurface $S$. Conservation of this charge, i.e., that it assumes 
constant values when computed on different spacelike hypersurfaces (corresponding 
to different times) is derived, as usual, when we integrate the conservation law 
$d J[\xi,\varepsilon] = 0$ over the $n$-volume domain with the boundary $S_1 + 
S_2 + T$, where $S_1$ and $S_2$ are $(n-1)$-dimensional spacelike hypersurfaces 
(which correspond to the arbitrary time values $t_1$ and $t_2$, respectively) 
and $T$ is a timelike surface that connects them. When the fields satisfy the 
``no-flux" boundary conditions such that $\int_T J = 0$, the charge (\ref{calq})
is constant. It is an interesting question whether the ``no-flux" condition might
be connected to the choice of $\xi$ as a Killing vector. We plan to study this 
hypothesis elsewhere.

We always assume that the fields satisfy the ``no-flux" condition, the explicit
form of which should be established on a case by case basis after the spacetime
dimension and the model Lagrangian is specified. One can check that these 
boundary conditions are fulfilled for all static and stationary configurations
which we consider in our subsequent computations. 

We now test the general formalism in the following concrete applications: 
(i) Einsteinian gravity with minimal coupling (general relativity in 4, 3, 
and 5 dimensions), (ii) a model with nonminimal coupling (Brans-Dicke 
theory), (iii) a higher-derivative gravity model. 

\section{Einstein(-Cartan) theory in any dimension} 

In $n$-dimensional spacetime,
the Hilbert-Einstein Lagrangian with cosmological constant $\lambda$ reads 
\begin{equation}
V = -{\frac 1{2\kappa_n}}\left(R^{\alpha\beta}\wedge\eta_{\alpha\beta} 
- 2\lambda\eta\right).\label{Vgr}
\end{equation}
For $n\geq 4$, the relativistic gravitational constant is $\kappa_n = 
2(n - 3)v_{n-1}G_n/c^3$, where $G_n$ is the Newtonian constant (the 
dimensionality of which depends on the dimension of space) and $v_d = 
2\pi^{d/2}/\Gamma(d/2)$ is the volume of a $(d-1)$-dimensional unit 
sphere. Then ${\cal Q}[\xi]=(1/2\kappa_n)\int_{\partial S}{}^\ast\left[
dk+\xi\rfloor(\vartheta^\lambda\wedge T_\lambda)\right]$, where 
$k:=\xi_\alpha\,\vartheta^\alpha$. This reduces to Komar's expression
for spinless matter or in vacuum, since then $T^\alpha=0$. 

{\it In 4 dimensions}, we consider the Lagrangian $V' = V + \alpha_0
\,d\Phi_P$ with (\ref{Vgr}) supplemented by a topological boundary 
term (cf. \cite{A00a}) given by the 3-form
\begin{equation}
\Phi_P = \eta_{\alpha\beta\mu\nu}\,\Gamma^{\alpha\beta}\wedge\left(
R^{\mu\nu}+ \frac{\hbox{$\scriptstyle{1}$}}{\hbox{$\scriptstyle{3}$}}
\,\Gamma^{\mu\lambda}\wedge\Gamma_\lambda{}^\nu\right).\label{CS3}
\end{equation}
The boundary term is needed to regularize the conserved quantities for 
asymptotically AdS configurations, which is achieved by choosing 
$\alpha_0 = 3/8\kappa_4\lambda$. 

Let us find the invariant charge for the Kerr-AdS solution: in the spherical
coordinates $(t,r,\theta,\varphi)$, the corresponding coframe reads 
\cite{OR06b}:
\begin{eqnarray}
\vartheta^{\hat 0} &=& \sqrt{\frac{\Delta}{\Sigma}}\left[ cdt
- a\Omega\sin^2\theta\,d\varphi\right],\label{cof0} \\
\vartheta^{\hat 1} &=& \sqrt{\frac{\Sigma}{\Delta}}\, dr,\quad
\vartheta^{\hat 2} = \sqrt{\frac{\Sigma}{f}}\, d\theta,\label{cof12} \\
\vartheta^{\hat 3} &=& \sqrt{\frac{f}{\Sigma}}\sin\theta\left[ -a\,cdt
+\Omega (r^2+a^2)\,d\varphi\right].\label{cof3}
\end{eqnarray}
Here $m = {G_4M}/{c^2}$, and the functions are defined by
\begin{eqnarray}
\Delta&:=& (r^2 + a^2)(1-\frac{\lambda}{3}\,r^2) - 2mr,\\
\Sigma&:=& r^2 + a^2\cos^2\theta,\\
f &:=& 1+\frac{\lambda}{3}\,a^2\cos^2\theta,\quad \Omega^{-1} 
:= 1+\frac{\lambda}{3}\,a^2. 
\end{eqnarray}
The invariant charges for this solution are then easily computed: 
\begin{equation}
{\cal Q}'[\partial_t]=\Omega Mc^2,\qquad {\cal Q}'[\partial_\varphi] 
=-\Omega^2Mca. 
\end{equation}
They coincide with the {\it noncovariant} charges found 
in \cite{A00a} for a particular choice of frame. In order to see this, one
must take into account that the coframe defined by (\ref{cof0})-(\ref{cof3})
differs from the one used in \cite{A00a}  by a factor $\Omega$ of the $dt$ 
component. In other words, the frame in \cite{A00a} corresponds to a change 
of time coordinate $t= \Omega t'$, so that ${\cal Q}'[\partial_t']=\Omega^2 
Mc^2$. One should always remember that the result of the use of the general 
formula (\ref{calq}) depends on the input, i.e., on the configuration of the
fields (metric and other), on the the integration domain in the integral, and 
on the vector field $\xi$. Careful application of this formula then reproduces
the same conserved charges as in \cite{GPP05}.

{\it Our general approach works in any dimension}: 
We consider now the models determined by the 
Lagrangian (\ref{Vgr}) for $n=3$ and $n=5$, respectively.

{\it 3D BTZ black hole:}  For the uncharged BTZ solution \cite{BTZ92}, 
we choose the frame: 
\begin{equation}
\vartheta^{\hat 0}=fc\,dt,\qquad \vartheta^{\hat 1}
= f^{-1}\,dr,\qquad \vartheta^{\hat 2} = rd\varphi - {\frac {Jc}{2r}}\,dt
\end{equation}
with $f = \sqrt{({J}/{2r})^2 - \lambda r^2 - m}$. The invariant charge 
${\cal Q}[\partial_t]$ formally diverges and regularization is needed.
This can be performed by the relocalization, $V\rightarrow V'=V+d\Phi$,
with the help of an appropriate boundary term. Explicitly, $\Phi =
-\eta_{\alpha\beta}\wedge\Delta\Gamma^{\alpha\beta}/2\kappa_3$ with 
$\Delta\Gamma_\alpha{}^\beta :=\Gamma_\alpha{}^\beta-\overline{\Gamma
}_\alpha{}^\beta$. The  ``background'' connection is chosen as a flat 
connection $\overline{\Gamma}_\alpha{}^\beta:=\left.\Gamma_\alpha
{}^\beta\right|_{m=J=0}$ with nontrivial components $\overline{\Gamma}
{}^{\hat 0\hat 1}=-\lambda rc\, dt$, $\overline{\Gamma}{}^{\hat 1\hat 2}
=-\sqrt{-\lambda}\, r\,d\varphi$. For $V'$ we then have 
\begin{equation}
{\cal Q}'[\xi]={\frac {1}{2\kappa_3}}\int_{\partial S}\eta_{\alpha\beta\lambda}
\,\xi^\alpha\Delta\Gamma^{\beta\lambda}.
\end{equation}
Direct computation for the BTZ solution yields
\begin{equation}
{\cal Q}'[\partial_t]= {\frac {\pi mc}{\kappa_3}},\qquad 
{\cal Q}'[\partial_\varphi]={\frac {\pi J}{\kappa_3}}.
\end{equation}
A similar but more involved derivation can be found in \cite{FFFR99}. 
Note that the relocalization above cancels the rotational contribution
[second term in (\ref{calq})] and replaces it with a translational one
[first term in (\ref{calq})]. This demonstrates the convenience of the 
general framework in which a Lagrangian may depend on all covariant 
geometrical objects, including the torsion. In this example, for 
the boundary term $d\Phi$ above, the derivative $H'_\alpha=-{\partial 
V'}/{\partial T^\alpha}\neq 0$ yields a nontrivial translational field
momentum despite the fact that torsion is absent, $T^\alpha=0$, ``on shell". 

{\it 5D Kerr solution} can be described, see for example \cite{GPP05},
by the line element 
\begin{eqnarray}
ds^2 &=& c^2dt^2 - {\frac \Sigma \Delta}\,dr^2 - {\frac {2m}\Sigma}\left(
cdt-a\sin^2\theta\,d\varphi-b\cos^2\theta\,d\psi\right)^2\nonumber\\
&& -\,\Sigma\, d\theta^2 - (r^2 + a^2)\sin^2\theta\,d\varphi^2 - 
(r^2+b^2)\cos^2\theta\,d\psi^2. 
\end{eqnarray}
Here $\Sigma=r^2 + a^2\cos^2\theta + b^2
\sin^2\theta$, $\Delta=(r^2+a^2)(r^2+b^2)/r^2-2m$, with $m=G_5M/c^2$ and 
$0<t<\infty$, $0<r<\infty$, $0<\theta<\pi/2$, $0<\varphi<2\pi$, $0<\psi<
2\pi$. We find the charges: 
\begin{equation}
{\cal Q}[\partial_t]=Mc^2/2,\qquad {\cal Q}[\partial_\varphi] =-Mca/2,
\qquad {\cal Q}[\partial_\psi]=-Mcb/2.
\end{equation}
The angular momenta obtained agree with the values given in \cite{GPP05}.
As for the total mass, its value is different from the value reported in
\cite{GPP05}, for example. However, this is the usual ``defect" of the Komar
charge which is easily repaired with the help of the appropriate total derivative
(boundary) term added to the Hilbert-Einstein Lagrangian, as was demonstrated
in \cite{Deruelle}.

\section{Brans-Dicke theory}

The Brans-Dicke theory is defined by the Lagrangian 
\begin{equation}\label{Vbd}
V=-\,{\frac 1{2\kappa_4}}\,(\phi\,\eta_{\alpha\beta}\wedge R^{\alpha\beta} 
+ \omega\phi^{-1}d\phi\wedge{}^\ast d\phi) - \psi_\alpha\wedge T^\alpha,
\end{equation}
where $\omega$ is a constant and the last term imposes the vanishing 
torsion condition. Then 
\begin{equation}
H_\alpha=\psi_\alpha,\qquad H_{\alpha\beta}=
{\frac {\phi}{2\kappa_4}}\,\eta_{\alpha\beta}.
\end{equation}
The field equation corresponding 
to variation w.r.t. the connection implies $\psi_\alpha=-e_\alpha\rfloor
{}^\ast d\phi/\kappa_4$. The spherically symmetric solution \cite{brans2}
is given in isotropic coordinates by the line element
\begin{equation}
ds^2=f^2c^2dt^2 - h^2(dr^2 + r^2d\Omega^2),
\end{equation}
with 
\begin{equation}
f = (1 - m/r)^{1/\mu}(1 + m/r)^{-1/\mu},\qquad
h = (1 + m/r)^2f^{\mu - \nu - 1}, 
\end{equation}
$m = G_4M/c^2$ and the scalar field $\phi = f^\nu$. The integration constants 
satisfy $\mu^2 = (1 + \nu)^2 - \nu(1 - \omega\nu)$. Then we find 
\begin{equation}
{\cal Q}[\partial_t] = Mc^2(1 - \nu)/\mu.
\end{equation} 
Our result agrees with the generalized Komar construction 
\cite{Pav} and differs from that of Hart \cite{Hart}. 

\section{Higher derivative gravity} 

Our general approach can also
be  applied to models with more nontrivial Lagrangians than (\ref{Vgr}) 
and (\ref{Vbd}). As a last example, let us now consider quadratic-curvature 
models in 4 dimensions. The Lagrangian 4-form of these models reads:
\begin{equation}
V = -\,{\frac 1{4\kappa_4}}\,R^{\alpha\beta}\wedge{}^\ast\left(\sum_{I=1}^6
\,b_I{}^{(I)}R^{\alpha\beta}\right) - \psi_\alpha\wedge T^\alpha.\label{Vq}
\end{equation}
The last term imposes the zero-torsion constraint.
As a result, the sum in the first term contains only three of the six 
irreducible pieces (see the definitions in \cite{OR06b}, e.g.): the Weyl 
2-form ${}^{(1)}R^{\alpha\beta}$, the traceless Ricci ${}^{(4)}R^{\alpha
\beta}$ and the curvature scalar ${}^{(6)}R^{\alpha\beta}$ piece. 
Accordingly, there are three coupling constants $b_1, b_4$ and $b_6$ 
(with dimension of length square) in the theory. In tensor language, the 
Lagrangian (\ref{Vq}) can be rewritten as 
\begin{equation}
V = -\,{\frac 1{4\kappa_4}}\left(\alpha R^2 + \beta {\rm Ric}_{\alpha\beta}
{\rm Ric}^{\alpha\beta} + \gamma R^{\alpha\beta}{}_{\mu\nu}R_{\alpha\beta}
{}^{\mu\nu}\right)\eta
\end{equation}
in terms of the curvature 
tensor components. The new coupling constants are related to the original 
ones via $\alpha = (2b_1 - 3b_4 + b_6)/12$, $\beta = b_4 - b_1$, $\gamma 
= b_1/2$. Sometimes, only the scalar square $R^2$ and the Ricci square 
${\rm Ric}^2_{\alpha\beta}$ terms are kept, whereas the total curvature 
quadratic term is ``removed" from the Lagrangian by making use of the 
Euler topological invariant $d\Phi_P$ (as done in \cite{Deser1}, 
for example). However, although the topological boundary term (\ref{CS3}) 
does not affect the field equations, it {\it does change} (as any other 
boundary term in the Lagrangian) the definition of field momenta and, 
hence, the conserved quantities.

For the Lagrangian (\ref{Vq}), we have 
\begin{equation}
H_\alpha = \psi_\alpha,\qquad H_{\alpha\beta} = {\frac 1{2\kappa_4}}
\,\sum_{I=1,4,6}\,b_I{}^\ast {}^{(I)}R_{\alpha\beta}. 
\end{equation}
The field equations yield for the Lagrange 
multiplier 
\begin{equation}
\psi_\alpha = 2e^\beta\rfloor DH_{\alpha\beta} - (1/2)
\vartheta_\alpha\wedge e_\mu\rfloor e_\nu\rfloor DH^{\mu\nu}.
\end{equation}
Substituting all this into ${\cal F}_\alpha = -DH_\alpha + E_\alpha 
= 0$ we obtain the system of ten fourth-order gravitational 
field equations. 

All the Einstein spaces, for which ${\rm Ric}_{\alpha\beta} = -\lambda
g_{\alpha\beta}$, are solutions of the fourth-order system for any 
constant $\lambda$. As an example, we choose again the Kerr-AdS spacetime
(\ref{cof0})-(\ref{cof3}). Like for the Einstein theory, a regularization 
is required for asymptotically nonflat solutions. For this purpose, we 
again use the relocalization $V' = V + \alpha_0d\Phi_P$, with the 
topological boundary term (\ref{CS3}). For the models (\ref{Vq}), this is
equivalent to a redefinition of the constants $b_1,b_4$ and $b_6$. We 
choose $\alpha_0=b_6/8\kappa_4$, and the direct computation for the 
Kerr-AdS solution then yields
\begin{equation}
{\cal Q}'[\partial_t] = \Lambda_0\Omega Mc^2,\qquad {\cal Q}'[\partial_\varphi]
= - \Lambda_0\Omega^2 Mca.
\end{equation}
Here the constant $\Lambda_0 := (b_6 - b_1)\lambda/3$ is dimensionless. This
qualitatively agrees (since $(b_6 - b_1)/3 = 4\alpha + \beta$) with the 
results of \cite{Deser1}, where a noninvariant definition of the total 
energy in quadratic-curvature theories was proposed. 

\section{Conclusion} 

We have presented a general definition of {\it invariant}
conserved currents and charges for gravity models with diffeomorphism and
local Lorentz symmetries. In our opinion, the advantage of our approach
is that everything is directly derived from the Lagrangian. The latter fixes
the physical laws that govern a system. On the contrary, it seems that the 
non-Noether currents do not have a direct physical meaning since they are
in general unrelated to the Lagrangian (and hence to the physical laws encoded 
in it). For that reason we excluded them from our analysis. We believe that 
our invariant Noether currents are physically meaningful quantities because: 
They i) are well defined for every Lagrangian, ii) satisfy the reasonable 
condition that current vanishes ${\cal J}=0$ for the trivial Lagrangian $V_0=0$, 
iii) lead to reasonable results when evaluated for known configurations. 
We have indeed verified that this approach yields 
satisfactory values of the total energy and angular momentum for 
asymptotically flat and asymptotically AdS solutions of the gravitational 
models in various dimensions, with various (minimal and nonminimal) coupling,
and for various (linear and quadratic) Lagrangians. We thus generalize and 
improve the results \cite{Wald,Komar,Trautman05,Benn,FFFR99,JS98,A00a}. 
We also confirm and strengthen the observation of \cite{A00a} that all 
locally AdS spacetimes have zero invariant charges for any $\xi$ both in
Einstein's gravity and in quadratic-curvature gravity, which implies 
degeneracy of the vacuum in these models. Among the number of possible 
further developments (currently under investigation), we mention the
interesting applications of this approach to supergravity, and to black 
hole thermodynamics along the lines of \cite{Wald}. 

{\bf Acknowledgments}. We thank J.G. Pereira for his 
hospitality at IFT-UNESP where this research was started. This work was 
partially supported by FAPESP and by DFG, project He~528/21-1 (for YNO) 
and by CNPq and by FONDECYT grant \#~1060939 (for GFR). We thank the 
anonymous referee for the useful discussion and advice.


\begin{thebibliography}{99}

\bibitem{footnote}
The volume $n$-form is
$\eta:=\vartheta^{\hat 0}\wedge\cdots\wedge\vartheta^{\hat n}$, and
$e_\alpha\rfloor\phi$ denotes the interior product of a form $\phi$ with the
vectors of the frame $e_\alpha$ dual to the coframe basis $\vartheta^\alpha$.
Latin and Greek indices label the 
coordinate and the local frame components, respectively (hats over an 
index denote individual components w.r.t. an anholonomic frame). For 
$p=0,\dots,n$, we define $\eta^{\alpha_1\cdots\alpha_p}:={}^\ast(
\vartheta^{\alpha_1}\wedge\cdots\wedge\vartheta^{\alpha_p})$ where 
${}^\ast$ is the Hodge operator. For other notation see \cite{OR06b}.

\bibitem{Wald}
V. Iyer and R.M. Wald, {\sl Phys. Rev.} {\bf D50} (1994) 846. 

\bibitem{Komar}
A. Komar, {\sl Phys. Rev.} {\bf 113} (1959) 934; 
{\sl Phys. Rev.} {\bf 127} (1962) 1411. 

\bibitem{Trautman05}
A. Trautman, {\it Einstein-Cartan theory}, in: 
{\sl Encyclopedia of Mathematical Physics},  Eds. J.P. Fran\c{c}oise, 
G.L. Naber, S.T. Tsou (Elsevier, Oxford, 2006) vol. 2,
p. 189. 

\bibitem{Benn}
I.M. Benn, {\sl Ann. Inst. H. Poincar\'e} {\bf A 37} (1982) 67. 

\bibitem{FFFR99} 
L. Fatibene, M. Ferraris, M. Francaviglia and M. Raiteri, 
{\sl Phys. Rev.} {\bf D 60} (1999) 124012; 
G. Allemandi, M. Francaviglia and M. Raiteri, 
{\sl Class. Quantum Grav.} {\bf 20} (2003) 5103.

\bibitem{JS98} 
B. Julia and S. Silva, 
{\sl Class. Quantum Grav.} {\bf 15} (1998) 2173.

\bibitem{A00a} 
R. Aros, M. Contreras, R. Olea, R. Troncoso, and J. Zanelli, 
{\sl Phys. Rev. Lett.} {\bf 84} (2000) 1647;
{\sl Phys. Rev.} {\bf D 62} (2000) 044002. 

\bibitem{Noether}
E. Noether, 
{\sl Nachr. Koenigl. Ges. Wiss. Goettinen, Math.-Phys. Kl.} (1918) 235-257.

\bibitem{Barnich}
G. Barnich and F. Brandt, 
{\sl Nucl. Phys.} {\bf B633} (2002) 3-82. 

\bibitem{Barnich2}
G. Barnich and G. Compere,
{\it Surface charge algebra in gauge theories and thermodynamic integrability},
arXiv:0708.2378. 

\bibitem{Wald2}
R.M. Wald, 
{\sl J. Math. Phys.} {\bf 31} (1990) 2378-2384. 

\bibitem{OR06b} 
Yu. N. Obukhov and G. F. Rubilar, {\sl Phys. Rev.} {\bf D74} (2006) 064002.

\bibitem{Lie}
General Lie derivatives are discussed in:
K. Yano, {\it The theory of Lie derivatives and its applications}, 
North-Holland, Amsterdam (1955);
Y. Kosmann, 
{\sl C.R. Acad. Sci. Paris} {\bf A262} (1966) 289; 394;
R.D. Hecht, F.W. Hehl, J.D. McCrea, E.W. Mielke, and Y. Ne'eman,  
{\sl Phys. Lett.} {\bf A172} (1992) 13;
E. Mielke, {\sl Phys. Rev.} {\bf D63} (2001) 044018.

\bibitem{BTZ92}
M.~Ba\~nados, C.~Teitelboim and J.~Zanelli,  
{\sl Phys. Rev. Lett.}  {\bf 69} (1992) 1849.

\bibitem{GPP05}
G.~W.~Gibbons, M.J. Perry and C.~N.~Pope,
{\sl Class.\ Quantum\ Grav.}  {\bf 22}, 1503 (2005). 

\bibitem{Deruelle}
N. Deruelle and J. Katz, 
{\it On the mass of a Kerr-anti-de Sitter spacetime in D dimensions},
{\sl Class. Quantum Grav.} {\bf 22} (2005) 421-424.

\bibitem{brans2}
C.H. Brans, {\sl Phys. Rev.} {\bf 125} (1962) 2194.

\bibitem{Pav}
R. Pavelle, {\sl Phys. Rev.} {\bf D8} (1973) 2369. 

\bibitem{Hart}
H.B. Hart, {\sl Phys. Rev.} {\bf D5} (1972) 1256; 
{\bf D11} (1975) 960.

\bibitem{Deser1}
S. Deser and B. Tekin, 
{\sl Phys. Rev. Lett.} {\bf 89} (2002) 101101. 


\end{thebibliography}
\end{document}